\documentclass[aps,prl,twocolumn,superscriptaddress]{revtex4}  

\usepackage{graphicx}
\usepackage{amsmath}
\usepackage{enumerate,amsthm,amssymb,color}
\usepackage{dsfont}
\usepackage{bbm}
\usepackage{bm}
\usepackage[french, english]{babel}
\usepackage{float}
\usepackage{newfloat}
\usepackage{upgreek}
\begin{document}
\title{Raman tailored photonic-crystal-fiber for telecom band photon-pair generation}

\author{M. Cordier}\affiliation{Laboratoire de Traitement et Communication de l'Information, T\'el\'ecom ParisTech, Universit\'e Paris-Saclay, 75013 Paris, France}
\author{A. Orieux}\affiliation{LIP6, CNRS, Universit\'e Pierre et Marie Curie, Sorbonne Universit\'es, 75005 Paris, France}\affiliation{IRIF, CNRS, Universit\'e Paris Diderot, Sorbonne Paris Cit\'e, 75013 Paris, France}
\author{R. Gabet}\affiliation{Laboratoire de Traitement et Communication de l'Information, T\'el\'ecom ParisTech, Universit\'e Paris-Saclay, 75013 Paris, France}
\author{T. Harl\'e}\affiliation{Laboratoire Charles Fabry, Institut d'Optique Graduate School, CNRS, Universit\'e Paris-Saclay, 91127 Palaiseau cedex, France}
\author{N. Dubreuil}\affiliation{Laboratoire Charles Fabry, Institut d'Optique Graduate School, CNRS, Universit\'e Paris-Saclay, 91127 Palaiseau cedex, France}
\author{E. Diamanti}\affiliation{LIP6, CNRS, Universit\'e Pierre et Marie Curie, Sorbonne Universit\'es, 75005 Paris, France}
\author{P. Delaye}\affiliation{Laboratoire Charles Fabry, Institut d'Optique Graduate School, CNRS, Universit\'e Paris-Saclay, 91127 Palaiseau cedex, France}
\author{I. Zaquine}\affiliation{Laboratoire de Traitement et Communication de l'Information, T\'el\'ecom ParisTech, Universit\'e Paris-Saclay, 75013 Paris, France}



      

\begin{abstract}
We report on the experimental characterization of a novel nonlinear liquid-filled hollow-core photonic-crystal fiber for the generation of photon pairs at telecommunication wavelength through spontaneous four-wave-mixing. We show that the optimization procedure in view of this application links the choice of the nonlinear liquid to the design parameters of the fiber, and we give an example of such an optimization at telecom wavelengths. Combining the modeling of the fiber and classical characterization techniques at these wavelengths, we identify, for the chosen fiber and liquid combination, spontaneous four-wave-mixing phase matching frequency ranges with no Raman scattering noise contamination. This is a first step toward obtaining a telecom band fibered photon-pair source with a high signal-to-noise ratio.
\end{abstract}



\maketitle


Photon pairs that can give rise either to entanglement or to heralded single photons are a basic resource for quantum information protocols for quantum communications or linear optics quantum computation. 
Recent achievements in entangled photon pair sources are mainly based on the second order nonlinear process of spontaneous parametric down-conversion (SPDC) in bulk crystals \cite{giustina2015significant,shalm2015strong}, and in waveguides \cite{alibart2016quantum}. The third order process of spontaneous four-wave mixing (SFWM) in fibers is however an interesting alternative towards integration in future quantum communication networks. Indeed, glass fibers are widely investigated for various nonlinear applications \cite{agrawal2007nonlinear}. They can exhibit very low absorption, strong confinement over long distance and high nonlinearities according to the chosen glass. Moreover, the various design parameters in microstructured fibers allow for a precise engineering of their propagation characteristics, including for instance dispersion. Among the third order nonlinear process associated with propagation in glass fibers, one can cite not only parametric process such as third harmonic generation or four-wave mixing, that can generate new frequencies in the medium if the phase-matching condition is satisfied, but also inelastic scattering process involving vibrational excitation modes of the medium such as Raman scattering. Due to the amorphous nature of glass, Raman scattering does not give rise to discrete lines but to a continuous broadband spectrum. This can be very useful for realizing broadband amplifiers \cite{bromage2004raman}, but quite deleterious for quantum optics applications such as photon pair generation where Raman photons give rise to uncorrelated-photon noise that cannot be filtered \cite{fulconis2005high,lin2007photon}. The resulting signal to noise ratio limitation is an important issue for the practical use of such sources for quantum information. 
 
In order to combine high nonlinearity with a discrete line Raman spectrum, the amorphous propagation medium can be replaced by a crystalline medium \cite{lin2006silicon,agha2012low}, a liquid \cite{barbier2015spontaneous}, or even a gas \cite{benabid2011linear}. In this last case Raman noise can be totally eliminated when using noble gases, and nonlinearity can be obtained at the cost of very high pressure conditions in the fiber.

A very practical solution is to use liquid-filled hollow-core  photonic-crystal fibers (LF-HC-PCF). They have been investigated either for designing Raman wavelength converters \cite{lebrun2007high,huy2010characterization,lebrun2010optical,lebrun2016efficient2} or conversely to show that the Raman lines can be avoided 
in order to produce high signal to noise ratio photon pairs through SFWM.
A near infrared LF-HC-PCF photon pair source was recently reported, exhibiting a three orders of magnitude suppression of the Raman noise \cite{barbier2015spontaneous}.

In this paper we detail the influence of the liquid on the guiding properties of the hollow-core photonic-crystal fiber and we show how the optimization of the microstructured fiber and liquid combination can provide low loss, Raman-free spectral zones in the selected telecom transmission band where photon pairs can be generated through FWM by satisfying the phase matching condition. Avoiding the spectral overlap of the two nonlinear processes can lead to a substantial increase of the signal to noise ratio (SNR) of the pair generation. We demonstrate the first nonlinear LF-HC-PCF with a transmission band and a zero dispersion wavelength (ZDW) in the telecom wavelength range as a first step towards a Raman-free photon pair source at such wavelengths. Prior to the description of the performed experimental characterizations, we give a modeling of the changes in the linear optical properties of the fiber induced by the liquid filling. These investigations show the strong links between fiber and liquid parameters that must be taken into account in order to satisfy all the requirements of a photon pair source.

The properties of the empty fiber and the choice of the liquid are related through the mapping of the transmission band of the filled fiber to the target SFWM phase matching range. 
The ZDW is the reference wavelength around which the phase matching condition can be satisfied. It should be ideally situated close to the middle of the transmission band in order to obtain a broad range of possible signal and idler wavelength pairs, which can be calculated from the dispersion properties of the filled fiber. The choice of the liquid is also constrained by its nonlinear coefficient, which determines the pair generation efficiency and its low absorption losses in the telecom range.
These LF-HC-PCF therefore provide a very versatile platform, allowing the engineering of linear and nonlinear properties through both fiber microstructure design and liquid optimization. The present work is based on the use of commercial fibers and focused on the choice of the best associated liquid.

On the practical side, both core and cladding are filled with the chosen liquid: by placing both ends of the fiber in tanks, the fiber automatically gets filled through capillarity. Liquids with relatively low viscosity (1 mPa.s) will therefore be selected in order to limit the fiber-filling process duration. The non-toxicity (or low toxicity) of the liquid is also taken into account. 
In this fully filled fiber, the photonic band gap guiding property is conserved despite the reduced index contrast between the glass part and the filled holes \cite{antonopoulos2006experimental}. Any characteristic wavelength $\lambda_{\text{empty}}$ of the transmission band of the empty fiber is shifted to $\lambda_{\text{filled}}$ in the following way, where the refractive index of a medium is noted $n_{\text{medium}}$  \cite{lebrun2016efficient2,antonopoulos2006experimental}:

\begin{equation}
\lambda_{\text{filled}} = \lambda_{\text{empty}}\sqrt{\frac{n^2_{\text{silica}}-n^2_{\text{liquid}}}{n^2_{\text{silica}}-n^2_{\text{air}}}}
\label{eq:shift}
\end{equation}

\begin{figure}[htbp]
\centering

\includegraphics[width=\linewidth]{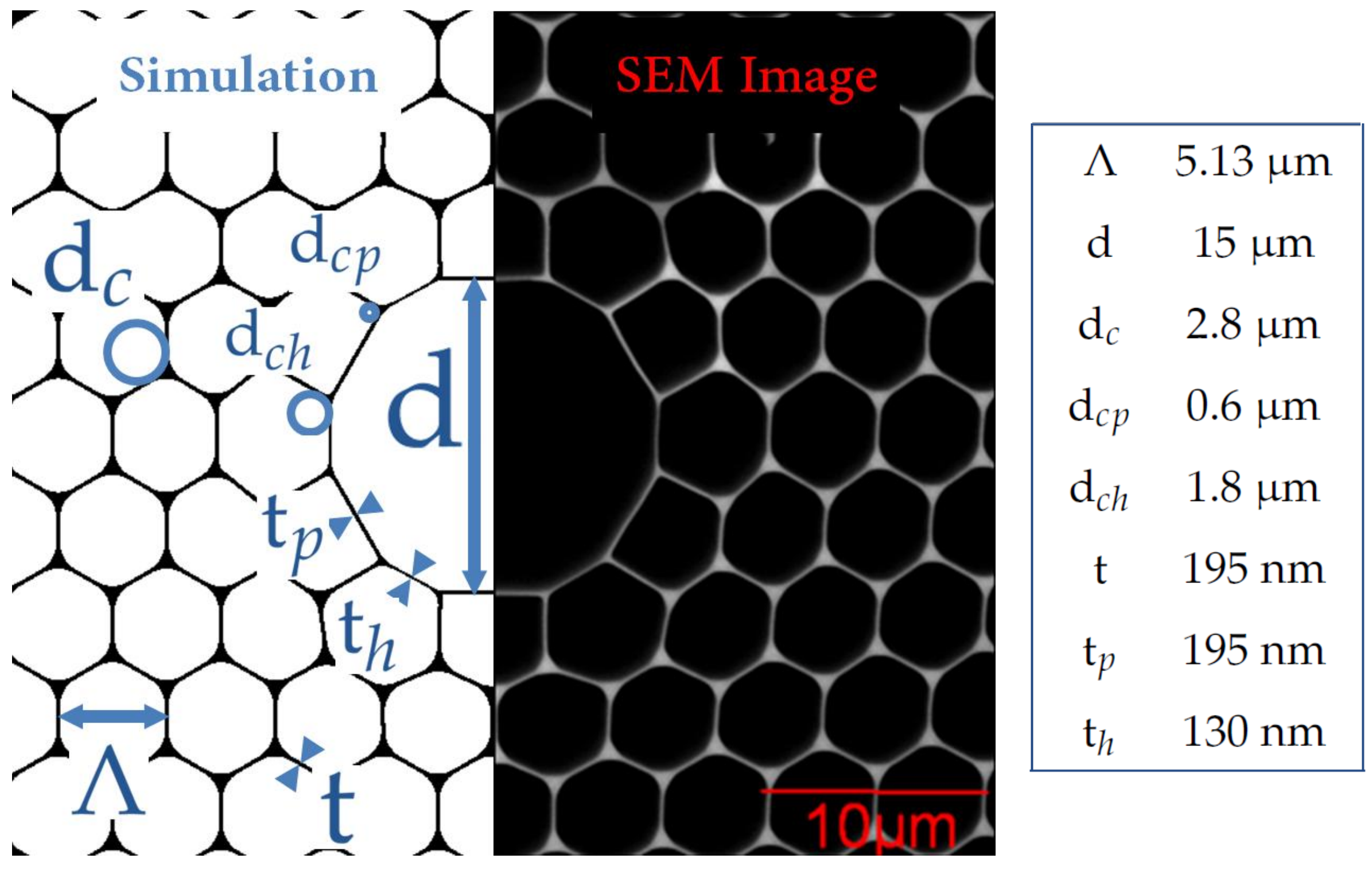}
\caption{Scanning electron microscope image of the microstructured region of fiber A (right part of the image) and the reconstructed geometry used for the simulation (left part), with the reconstruction parameters in the table : pitch $\Lambda$ , core diameter $d$, diameter of the round corners of the cladding hexagons $d_c$, diameters of the cladding pentagons $d_{cp}$ and hexagons $d_{ch}$ neighbouring the fiber core, cladding thickness $t$ and cladding thickness around the fiber core $t_p$ and $t_h$.}
\label{fig:SEB}
\end{figure}

To operate nearby the telecom wavelength 1.55 $\upmu $m, the empty PCF transmission must therefore be centered well beyond 1.55 $\upmu $m. 
We have chosen two commercial fibers (NKT Photonics) with transmission bands centered around 2.3 $\upmu $m (called Fiber A) and 2 $\upmu $m (called Fiber B). Eq. (\ref{eq:shift}) provides a first estimation of the liquid refractive index range (1.25-1.29) in order to shift the transmission window to the telecom band. Fluorocarbon liquids can be used with the further advantage of a low absorption in this range. In order to predict precisely the dispersion properties of the filled fiber, we investigated numerically the impact of the filling on the transmission window for various combinations of fibers and liquids. The finite difference frequency domain (FDFD) simulations \cite{fallahkhair2008vector} are based on a reconstructed fiber geometry. The parameters are directly extracted from SEM images of the fiber (see Fig. \ref{fig:SEB}).
It is noteworthy that the parameters of pentagons and hexagons around the core are slightly different from the rest of the cladding as shown in  \cite{aghaie2013experimental}. The limited precision in fiber parameter extraction and the uncertainty on the exact refractive index value of the liquid at telecom wavelength contribute to an error on the absolute position of the ZDW. We can estimate an upper bound of $\pm 30 \text{nm}$ to this error that mostly corresponds to a global shift of the group velocity dispersion curve with no impact on the curvature. Moreover, in the case of fluorocarbon liquids selected in this work, we have checked that the liquid dispersion can be neglected compared to the waveguide dispersion.

\begin{figure}[htbp]
\centering
\includegraphics[width=\linewidth]{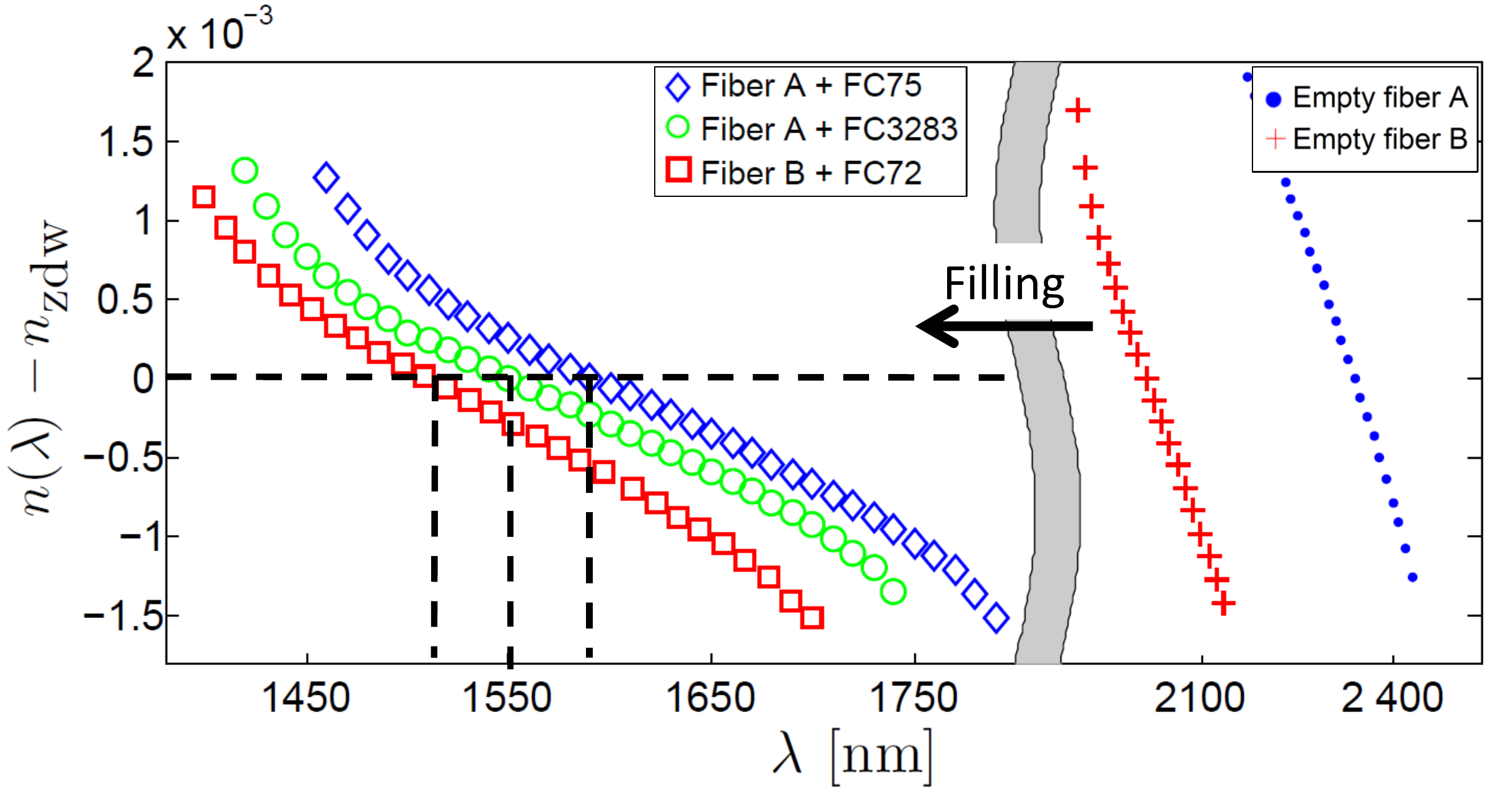}
\caption{Calculated effective index of the fundamental mode as a function of wavelength. Note that the subtracted $n_{\text{zdw}}$ has a different value for the three curves.}
\label{fig:index}
\end{figure}

Figure \ref{fig:index} shows the FDFD simulation results of the three tested combinations. We plot a relative refractive index, subtracting $n_{\text{zdw}}$ (index value at the ZDW), to allow for an easier comparison of the zero dispersion wavelengths in the three cases. 

\begin{figure}[htbp]
\centering
\includegraphics[width=\linewidth]{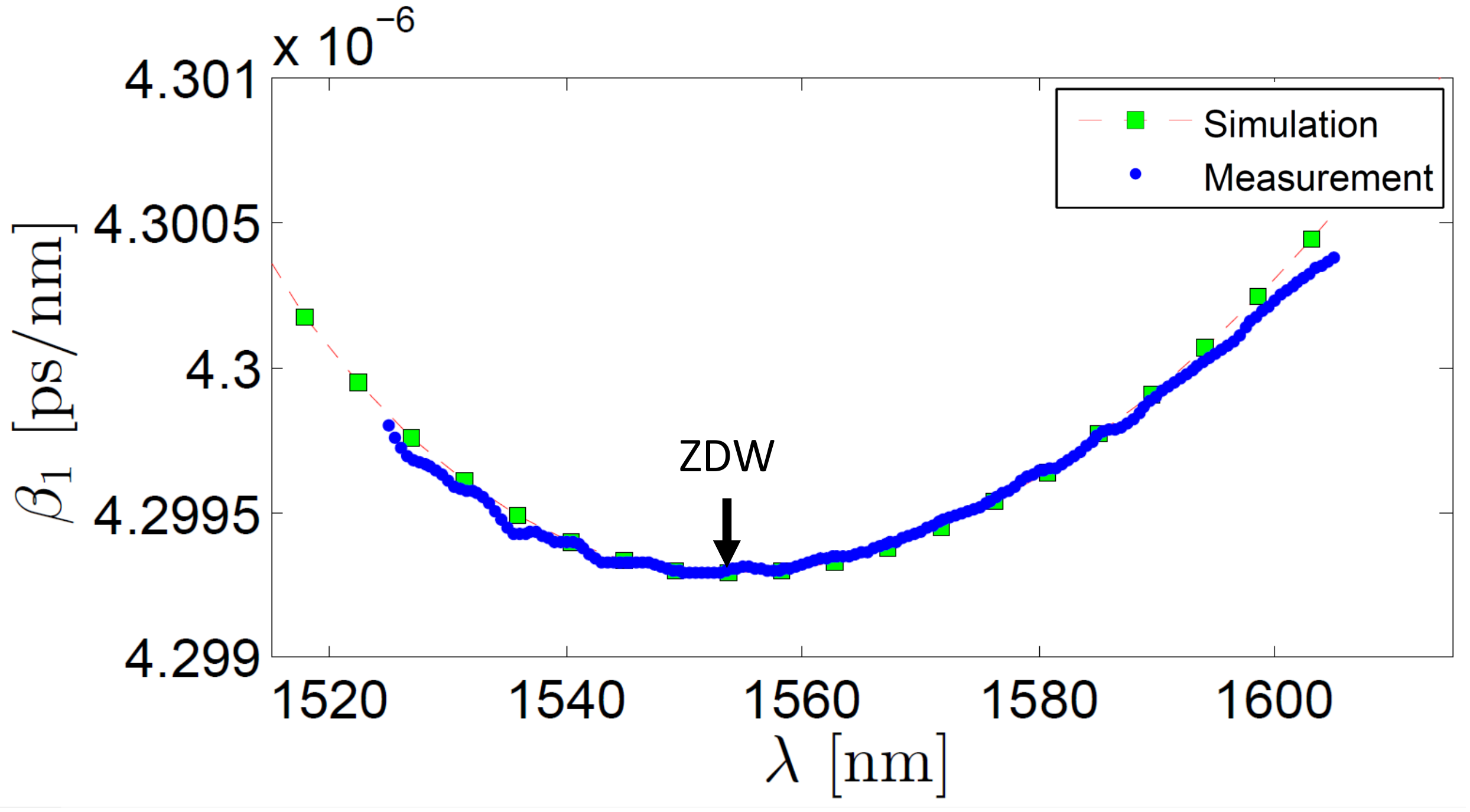}
\caption{Inverse group velocity of the fundamental mode for the Fiber A / FC3283 combination (i) measured with OLCI (Blue dots) (ii) calculated with the FDFD simulation (green squares). The index of the liquid is adjusted to 1.283 instead of the reported value of 1.281 measured in the visible range \cite{hasi2008investigation}.}
\label{fig:beta1}
\end{figure}

The dispersion of the three simulated fibers has been characterized by Optical Low-Coherence Interferometry (OLCI): the interferometer experimental setup is described in Fig. 1 of Ref \cite{gabet2009versatile}. The measured interferogram allows the calculation of the complex transmittivity from which the group-velocity dispersion (GVD) of each transmitted mode is derived. Table \ref{Table:combination} summarizes the three tested fiber characteristics. 
Only the fiber filled with FC3283 (perfluorotripropylamine, $\text{C}_9\text{F}_{21}\text{N}$) with a refractive index of 1.28 exhibits a ZDW close to $1.55\upmu \text{m}$ thus satisfying the requirement for a photon pair source in the C-band. For this fiber/ liquid combination the measured inverse group-velocity $\beta_1$ of the fundamental mode is shown in Figure \ref{fig:beta1} and compared with the simulation. The measured ZDW is $1.552\upmu \text{m} \pm 3 \text{nm} $. The FDFD simulated curve is in perfect agreement with the OLCI measurement of group velocity dispersion. This technique also gives access to the dispersion of higher propagation modes which can highlight coupling between different guided modes. We are especially interested in trying to avoid surface modes that can induce anti-crossing with the fundamental mode \cite{west2004surface}.
In such a case, the guided mode would then be a combination of the fundamental mode and a surface mode, with higher overlap with the silica structure inducing therefore more losses, dispersion and Raman noise.
We have checked that the chosen fiber-liquid combination does not exhibit such an anti-crossing in a 100 nm wide window around the ZDW where the photon pairs will be generated. \\
 \begin{table}[htb]
 \centering \caption{Optical properties for different fiber/liquid combinations. The 3M references FCxxx have been used for the fluorocarbon liquids. The NKT reference HCxxxx gives an approximate idea of the position of the transmission band of the empty fiber and the ZDW of the filled fiber has been measured with OLCI. The configuration of Ref. \cite{barbier2015spontaneous} is also included for comparison.}
\begin{tabular}{|c|ccc|}
\hline
\small
Fiber & Liquid & $n_{\text{liquid}} $&  ZDW (nm) \\
\hline \hline
A \small{(HC2300)}& \small{FC3283} & 1.28 & 1552  \\
A \small{(HC2300)}& \small{FC75} & 1.27 & 1592  \\
B \small{(HC2000)}& \small{FC72} & 1.25 & $\approx 1510$   \\
\cite{barbier2015spontaneous} \small{(HC1550PM)}& \small{Acetone d6} & 1.36 & 896   \\ 
\hline 
\end{tabular}
 \label{Table:combination}
    \end{table}
\normalsize

\begin{figure}[htbp]
\centering
\includegraphics[width=\linewidth]{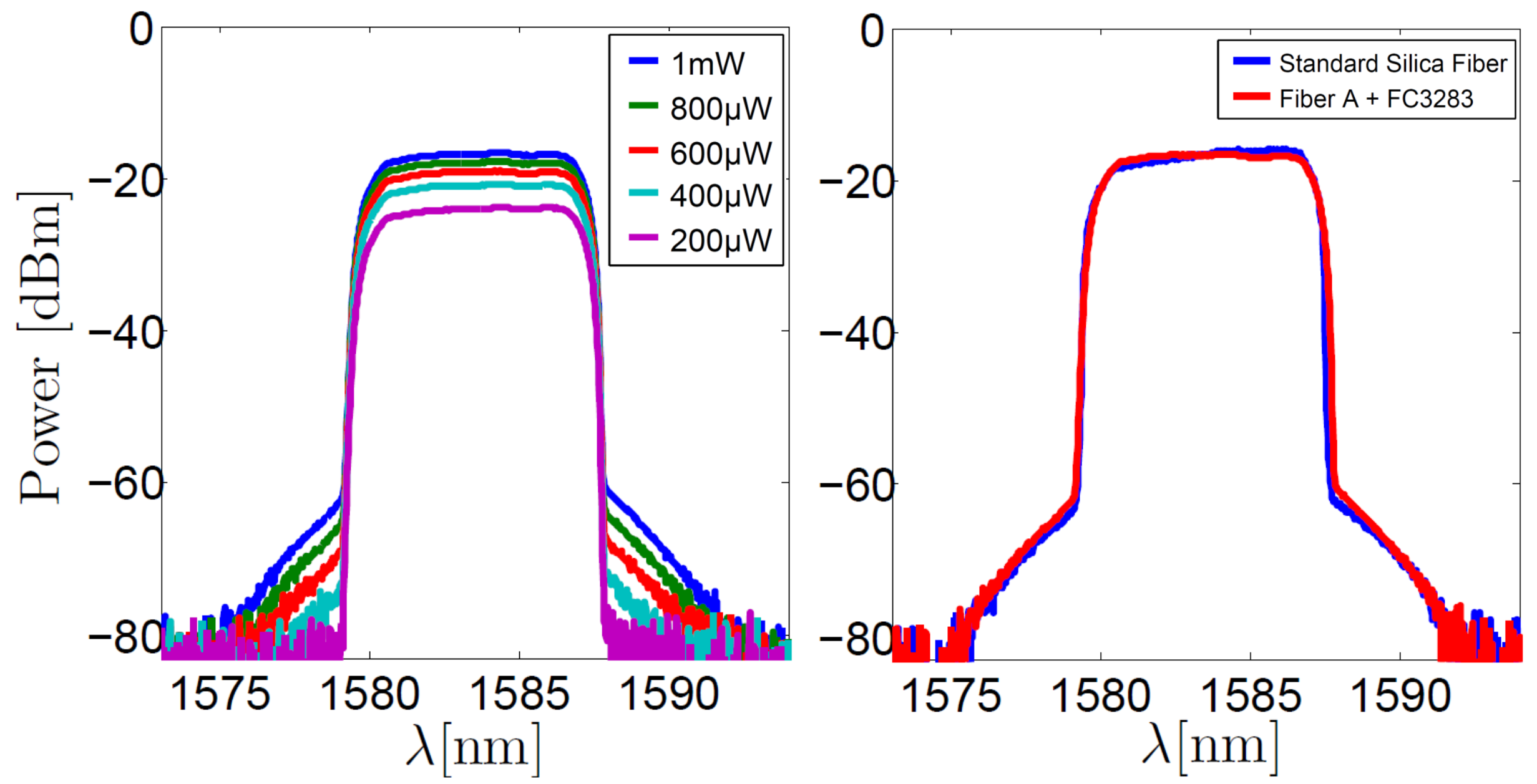}
\caption{The laser delivers 1.1 ps pulses (close to Fourier transform) with a 50MHz repetition rate ; the peak power is 18W for 1mW average power; Left : Spectra of the square wave pulse generator transmitted by the fiber for different levels of injected power. Right : Comparison between the fiber under study and a standard polarization maintaining fiber.}
\label{fig:n2}
\end{figure}

The nonlinearity of the LF-HC-PCF is also an important parameter for the optimization of the signal to noise ratio of the photon pair generation. Because we use the fundamental propagation mode with a low overlap with the silica structure, it is determined solely by the nonlinear index of the liquid. We measure the nonlinear index of the FC3283 through self-phase modulation broadening of square spectrum pulses, using the setup described in \cite{serna2015enhanced}. The transmission spectra measured for increasing incident power are shown in the left graph of Figure  \ref{fig:n2}. Whereas the spectrum at low power exhibits a square shape, a symmetric spectral broadening is observed for increasing power, characterizing the self-phase modulation process (SPM) induced by optical Kerr effect. By comparing with simulated spectra, the nonlinear phase shift $\Phi_{\text{NL}} = n_2 \frac{2 \pi L}{A_{\text{eff}} \lambda } P_{\text{peak}}$ can be derived from the measured spectral broadening, with $\text{n}_2 $ the nonlinear refractive index,  $L$ the fiber length, $A_{\text{eff}}$ the effective area, $\lambda$ the wavelength and $P_{\text{peak}}$ the peak power injected in the fiber mode.
The square spectral shape of the pump pulses offers an improved sensitivity in measuring low nonlinear phase shifts. The lowest nonlinear phase shift observed at 400 µW (see the pedestals on the spectra) is about 4 mrad. A nonlinear phase shift of $\Phi_{\text{NL}} = 10 \text{mrad}$ is obtained for an average power of 1mW in the LF-HC-PCF. This measured value is of the order of the one obtained with a polarization maintaining silica fiber of the same length (Fig. \ref{fig:n2}). Taking into account the experimental parameters and the characteristics of the fiber ($A_{\text{eff}} =95\upmu \text{m}^2$, $L=1\text{m}$) , this leads to an estimated nonlinearity of the chosen liquid FC3283 $n_2=1.1$x$10^{-20}\text{m}^2/\text{W}$. This nonlinear coefficient, together with experimental parameters depending on the chosen setup will determine the conversion efficiency of the SFWM process in this LF-HC-PCF. 

It is noteworthy that nonlinearities as large as two orders of magnitude higher than silica can be obtained with other liquids, such as carbon disulfide (hence giving rise to a four orders of magnitude increase in the SFWM efficiency). These high nonlinearities are often associated to a high refractive index, thus requiring the use of fibers with glass index higher than silica. Higher index glass HC-PCF have been reported \cite{desevedavy2010chalcogenide,jiang2011single}, but they remain more difficult to obtain than silica HC-PCF.

\begin{figure}[htbp]
\centering
\includegraphics[width=\linewidth]{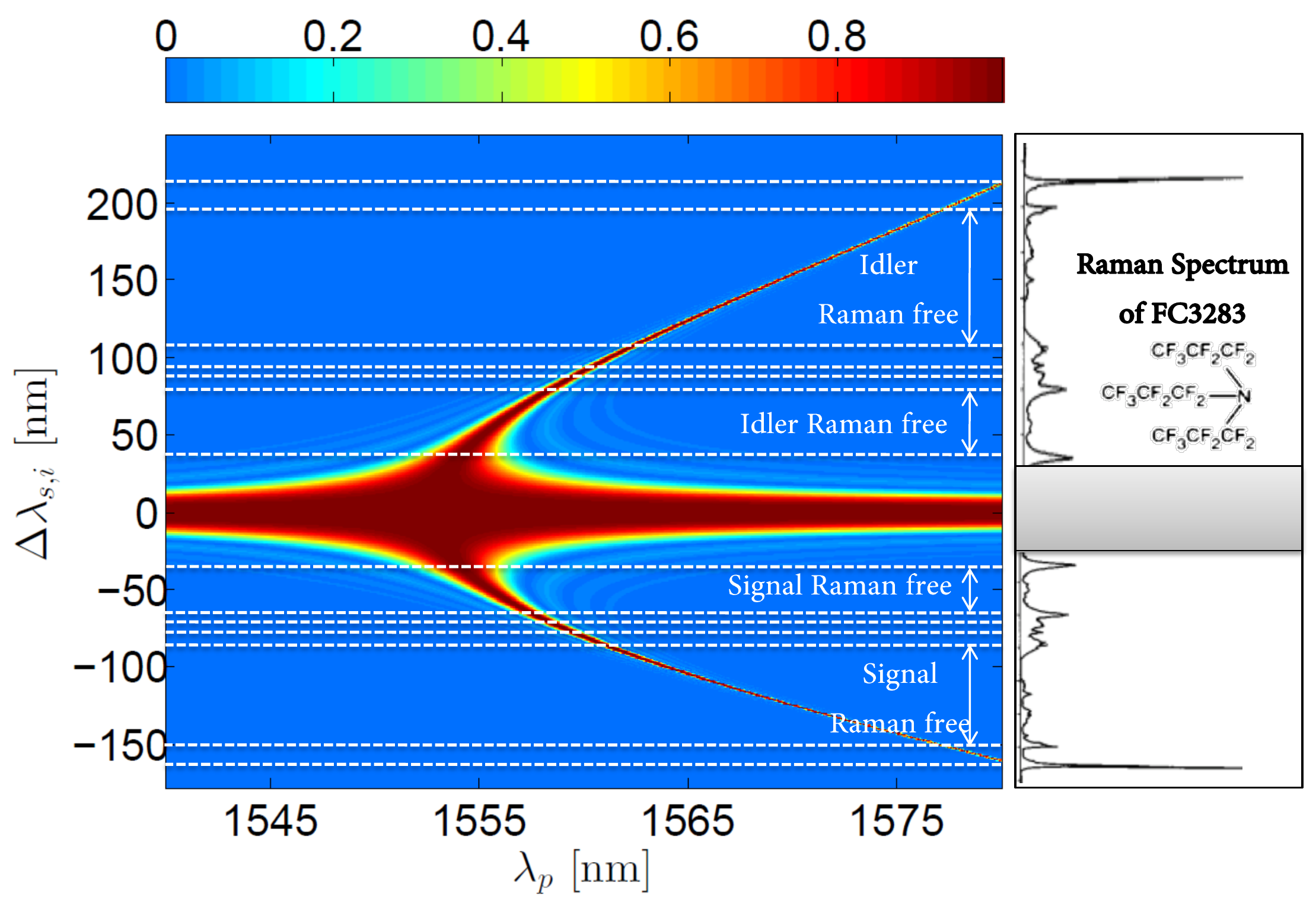}
\caption{Left: Calculated probability density spectrum $F$ of emission of the pairs through SFWM and Raman lines (white dashed lines) as a function of the pump wavelength $\lambda_p$ and the relative signal or idler wavelength $\Delta \lambda_{s,i}=\lambda_{s,i}-\lambda_p$; Right: Raman spectrum of the chosen fluorocarbon liquid: perfluorotripropylamine, $\text{C}_9\text{F}_{21}\text{N}, \, \text{FC}3283$, reproduced with permission from \cite{gorelik1999raman2} }
\label{fig:phase_matching}
\end{figure}

From the classical linear and nonlinear characterization presented in the previous part, and from the Raman spectrum of the FC3283 (see Fig. \ref{fig:phase_matching}), we can predict the SFWM phase-matching curve of the fiber and identify Raman-free zones for the photon pair generation. In the case of SFWM with degenerate pumps, signal and idler frequencies are constrained by the energy : $2 \omega_p = \omega_s + \omega_i$, and momentum conservation : $2 \beta(\omega_p) = \beta(\omega_s) + \beta(\omega_i)$ where $\omega_{p,s,i}$  and  $\beta(\omega_{p,s,i})$ respectively refer to the frequencies and wavevector of the pump, signal and idler photons \cite{agrawal2007nonlinear}. 
We define the phase-matching function $F(\omega_{p0},\Delta\omega)$ as: 
\begin{align}
F\!= \!\text{sinc}^2((2 \beta(\omega_{p0}) \!- \!\beta(\omega_{p0} \!+ \!\Delta\omega) \!- \! \beta(\omega_{p0} \!- \!\Delta\omega))  \, \frac{L}{2} ) 
\end{align} 
with $L$ the fiber length and $\Delta \omega = \omega_{p0} - \omega_i = \omega_s - \omega_{p0}  $ the frequency difference between pump and idler or signal and pump.

The phase matching function F, calculated in the case of a strict energy conservation for a pump frequency  $\omega_{p0}$, can also be seen as a normalized probability density spectrum of photon pair emission in the LF-HC-PCF. Its dependence with pump wavelength and relative signal or idler wavelength is shown in Fig. \ref{fig:phase_matching}. This two dimensional plot allows to choose the experimental parameters : in the central red zone corresponding to degeneracy of the signal and idler wavelengths, it is difficult to separate the photon pairs from the pump. We are rather interested in the quasi-parabolic parts on both sides with a large wavelength separation between signal and idler photons. The superimposition of the Raman lines of the liquid on this map shows the pump spectral ranges for which signal and idler photons can be generated away from the Raman noise. 

We have shown that the choice of the liquid allows to engineer both linear and non linear properties of a LF-HC-PCF and more specifically, the ZDW position, the nonlinear coefficient and the Raman free zones. In this way, the phase-matching wavelength range of the four wave mixing process and the photon pair generation efficiency can be optimized with respect to the Raman line position.
For a given commercial fiber, we have identified at least one liquid that satisfies all requirements toward the realization of a LF-HC-PCF Raman free photon pair source, with a ZDW in the desired telecom range, and a nonlinear coefficient comparable to what is obtained with silica. 
All the elements of a high SNR fibered photon pair source at telecom wavelength, fully operating at room temperature are therefore gathered.
Moreover the engineering of nonlinearity, spectral correlation of the signal and idler photons could be further improved, using a fiber especially designed for this purpose.
With the possibility to design specific fibers, and to use various glass and liquid media, the investigation of such a versatile fiber medium at telecom wavelength can give rise not only to high quality sources but also to a variety of devices and therefore have a large impact on quantum communication network development.

\section*{Acknowlegments}
We thank Stephan Suffit in Laboratoire Matériaux et Phénomènes Quantiques (CNRS, Université Paris 7) for the SEM images of the fiber. We thank the 3M company for the free sample of FC3283.
\section*{Funding Information}
This work is supported by the "IDI 2016" project funded by the IDEX Paris-Saclay, ANR-11-IDEX-0003-02

\bibliographystyle{osajnl}
\bibliography{sample}

\begin{thebibliography}{10}
\newcommand{\enquote}[1]{``#1''}

\bibitem{giustina2015significant}
M.~Giustina, M.~A. Versteegh, S.~Wengerowsky, J.~Handsteiner, A.~Hochrainer,
  K.~Phelan, F.~Steinlechner, J.~Kofler, J.-{\AA}. Larsson, C.~Abell{\'a}n,
  W.~Amaya, V.~Pruneri, M.~W. Mitchell, J.~B.~T. Gerrits, A.~E. Lita, L.~K.
  Shalm, S.~W. Nam, T.~Scheidl, R.~Ursin, B.~Wittmann, and A.~Zeilinger,
  \enquote{Significant-loophole-free test of bell’s theorem with entangled
  photons,} Physical review letters \textbf{115}, 250401 (2015).

\bibitem{shalm2015strong}
L.~K. Shalm, E.~Meyer-Scott, B.~G. Christensen, P.~Bierhorst, M.~A. Wayne,
  M.~J. Stevens, T.~Gerrits, S.~Glancy, D.~R. Hamel, M.~S. Allman, K.~J.
  Coakley, S.~D. Dyer, C.~Hodge, A.~E. Lita, V.~B. Verma, C.~Lambrocco,
  E.~Tortorici, A.~L. Migdall, Y.~Zhang, D.~R. Kumor, W.~H. Farr, F.~Marsili,
  M.~D. Shaw, J.~A. Stern, C.~Abell\'an, W.~Amaya, V.~Pruneri, T.~Jennewein,
  M.~W. Mitchell, P.~G. Kwiat, J.~C. Bienfang, R.~P. Mirin, E.~Knill, and S.~W.
  Nam, \enquote{Strong loophole-free test of local realism,} Phys. Rev. Lett.
  \textbf{115}, 250402 (2015).

\bibitem{alibart2016quantum}
O.~Alibart, V.~D’Auria, M.~De~Micheli, F.~Doutre, F.~Kaiser, L.~Labont{\'e},
  T.~Lunghi, E.~Picholle, and S.~Tanzilli, \enquote{Quantum photonics at
  telecom wavelengths based on lithium niobate waveguides,} Journal of Optics
  \textbf{18}, 104001 (2016).

\bibitem{agrawal2007nonlinear}
G.~P. Agrawal, \emph{Nonlinear fiber optics} (Academic press, 2007).

\bibitem{bromage2004raman}
J.~Bromage, \enquote{Raman amplification for fiber communications systems,}
  Journal of Lightwave Technology \textbf{22}, 79--93 (2004).

\bibitem{fulconis2005high}
J.~Fulconis, O.~Alibart, W.~Wadsworth, P.~S.~J. Russell, and J.~Rarity,
  \enquote{High brightness single mode source of correlated photon pairs using
  a photonic crystal fiber,} Optics Express \textbf{13}, 7572--7582 (2005).

\bibitem{lin2007photon}
Q.~Lin, F.~Yaman, and G.~P. Agrawal, \enquote{Photon-pair generation in optical
  fibers through four-wave mixing: Role of raman scattering and pump
  polarization,} Physical Review A \textbf{75}, 023803 (2007).

\bibitem{lin2006silicon}
Q.~Lin and G.~P. Agrawal, \enquote{Silicon waveguides for creating
  quantum-correlated photon pairs,} Optics letters \textbf{31}, 3140--3142
  (2006).

\bibitem{agha2012low}
I.~Agha, M.~Davan{\c{c}}o, B.~Thurston, and K.~Srinivasan, \enquote{Low-noise
  chip-based frequency conversion by four-wave-mixing bragg scattering in
  {S}i{N}x waveguides,} Optics letters \textbf{37}, 2997--2999 (2012).

\bibitem{barbier2015spontaneous}
M.~Barbier, I.~Zaquine, and P.~Delaye, \enquote{Spontaneous four-wave mixing in
  liquid-core fibers: towards fibered raman-free correlated photon sources,}
  New Journal of Physics \textbf{17}, 053031 (2015).

\bibitem{benabid2011linear}
F.~Benabid and P.~Roberts, \enquote{Linear and nonlinear optical properties of
  hollow core photonic crystal fiber,} Journal of Modern Optics \textbf{58},
  87--124 (2011).

\bibitem{lebrun2007high}
S.~Lebrun, P.~Delaye, R.~Frey, and G.~Roosen, \enquote{High-efficiency
  single-mode raman generation in a liquid-filled photonic bandgap fiber,}
  Optics letters \textbf{32}, 337--339 (2007).

\bibitem{huy2010characterization}
M.~C. Phan~Huy, A.~Baron, S.~Lebrun, R.~Frey, and P.~Delaye,
  \enquote{Characterization of self-phase modulation in liquid filled hollow
  core photonic bandgap fibers,} JOSA B \textbf{27}, 1886--1893 (2010).

\bibitem{lebrun2010optical}
S.~Lebrun, C.~Buy, P.~Delaye, R.~Frey, G.~Pauliat, and G.~Roosen,
  \enquote{Optical characterizations of a raman generator based on a hollow
  core photonic crystal fiber filled with a liquid,} Journal of Nonlinear
  Optical Physics \& Materials \textbf{19}, 101--109 (2010).

\bibitem{lebrun2016efficient2}
S.~Lebrun, M.~C. Phan~Huy, P.~Delaye, and G.~Pauliat, \enquote{Efficient
  stimulated raman scattering in hybrid liquid-silica fibers for wavelength
  conversion,} Proceeding SPIE \textbf{10021}, 1002104--1002104 (2016).

\bibitem{antonopoulos2006experimental}
G.~Antonopoulos, F.~Benabid, T.~Birks, D.~Bird, J.~Knight, and P.~S.~J.
  Russell, \enquote{Experimental demonstration of the frequency shift of
  bandgaps in photonic crystal fibers due to refractive index scaling,} Optics
  Express \textbf{14}, 3000--3006 (2006).

\bibitem{fallahkhair2008vector}
A.~B. Fallahkhair, K.~S. Li, and T.~E. Murphy, \enquote{Vector finite
  difference modesolver for anisotropic dielectric waveguides,} Journal of
  Lightwave Technology \textbf{26}, 1423--1431 (2008).

\bibitem{aghaie2013experimental}
K.~Z. Aghaie, M.~J. Digonnet, and S.~Fan, \enquote{Experimental assessment of
  the accuracy of an advanced photonic-bandgap-fiber model,} Journal of
  Lightwave Technology \textbf{31}, 1015--1022 (2013).

\bibitem{gabet2009versatile}
R.~Gabet, P.~Hamel, Y.~Jaou{\"e}n, A.-F. Obaton, V.~Lanticq, and G.~Debarge,
  \enquote{Versatile characterization of specialty fibers using the
  phase-sensitive optical low-coherence reflectometry technique,} Journal of
  Lightwave Technology \textbf{27}, 3021--3033 (2009).

\bibitem{west2004surface}
J.~A. West, C.~M. Smith, N.~F. Borrelli, D.~C. Allan, and K.~W. Koch,
  \enquote{Surface modes in air-core photonic band-gap fibers,} Optics Express
  \textbf{12}, 1485--1496 (2004).

\bibitem{hasi2008investigation}
W.~Hasi, Z.~Lu, S.~Gong, S.~Liu, Q.~Li, and W.~He, \enquote{Investigation of
  stimulated brillouin scattering media perfluoro-compound and
  perfluoropolyether with a low absorption coefficient and high power-load
  ability,} Applied optics \textbf{47}, 1010--1014 (2008).

\bibitem{serna2015enhanced}
S.~Serna, J.~Oden, M.~Hanna, C.~Caer, X.~Le~Roux, C.~Sauvan, P.~Delaye,
  E.~Cassan, and N.~Dubreuil, \enquote{Enhanced nonlinear interaction in a
  microcavity under coherent excitation,} Optics express \textbf{23},
  29964--29977 (2015).

\bibitem{desevedavy2010chalcogenide}
F.~D{\'e}s{\'e}v{\'e}davy, G.~Renversez, J.~Troles, P.~Houizot, L.~Brilland,
  I.~Vasilief, Q.~Coulombier, N.~Traynor, F.~Smektala, and J.-L. Adam,
  \enquote{Chalcogenide glass hollow core photonic crystal fibers,} Optical
  Materials \textbf{32}, 1532--1539 (2010).

\bibitem{jiang2011single}
X.~Jiang, T.~Euser, A.~Abdolvand, F.~Babic, F.~Tani, N.~Joly, J.~Travers, and
  P.~S.~J. Russell, \enquote{Single-mode hollow-core photonic crystal fiber
  made from soft glass,} Optics express \textbf{19}, 15438--15444 (2011).

\bibitem{gorelik1999raman2}
V.~S. Gorelik, A.~V. Chervyakov, L.~I. Zlobina, and O.~N. Sharts,
  \enquote{Raman and fluorescence spectra of fluoro-organic compounds,}
  Proceeding SPIE \textbf{3855}, 16--27 (1999).

\end{thebibliography}

\bigskip
\end{document}